# Gas-induced selective re-orientation of Au-Cu nanoparticles on TiO$_2$ (110)

Axel WILSON[a,c,*], Aude BAILLY[d], Romain BERNARD[a], Yves BORENSZTEIN[a], Alessandro COATI[c], Bernard CROSET[a], Hervé CRUGUEL[a], Ahmed NAITABDI[e], Mathieu SILLY[c], Marie-Claire SAINT-LAGER[d], Alina VLAD[c], Nadine WITKOWSKI[a], Yves GARREAU[b,c], Geoffroy PREVOT[a,*]

a Sorbonne Université, CNRS, Institut des NanoSciences de Paris, UMR 7588, F-75005, Paris, France

b Université Paris Diderot, Sorbonne-Paris-Cité, MPQ, UMR 7162 CNRS, Bâtiment Condorcet, Case 7021, 75205 Paris CEDEX 13, France

c Synchrotron SOLEIL, L'Orme des Merisiers Saint-Aubin - BP 48 91192 Gif-sur-Yvette CEDEX, France

d Institut Néel, CNRS et Université Joseph Fourier, BP 166, F-38042 Grenoble Cedex 9, France

e Sorbonne Université, CNRS, Laboratoire de Chimie Physique Matière et Rayonnement, 4, place Jussieu 75005 Paris

* Corresponding authors: axel.wilson@diamond.ac.uk, prevot@insp.jussieu.fr

**Abstract**

Au-Cu bimetallic nanoparticles (NPs) grown on TiO$_2$(110) have been followed *in-situ* using grazing incidence x-ray diffraction and x-ray photoemission spectroscopy from their synthesis to their exposure to a CO/O$_2$ mixture at low pressure (P < 10$^{-5}$ mbar) and at different temperatures (300 K–470 K). As-prepared samples are composed of two types of alloyed NPs: randomly oriented and epitaxial NPs. Whereas the introduction of CO has no effect on the structure of the NPs, an O$_2$ introduction triggers a Cu surface segregation phenomenon resulting in the formation of a Cu$_2$O shell reducible by annealing the sample over 430 K. A selective re-orientation of the nanoparticles, induced by the exposure to a CO/O$_2$ mixture, is observed where the randomly oriented NPs take advantage of the mobility induced by the Cu segregation to re-orient their Au-rich core relatively to the TiO$_2$(110) substrate following specifically the orientation ((111)$_{NPs}$//(110)$_{TiO2}$) when others epitaxial relationships were observed on the as-prepared sample.

**I) Introduction**

Supported metallic nanoparticles (NPs) have a prominent place in the field of catalysis. Haruta's discovery showing that Au NPs activity exceptionally improves when their size is reduced down to a few nm[1,2,3] triggered an intense and fruitful research activity in catalysis.[4,5,6,7,8] This activity was further stimulated by studies showing that the addition of a second metal could improve the catalytic properties of NPs by either promoting their activity and selectivity or increasing their stability.[9,10,11,12,13]

The enhancement of the activity and/or the stability of the NPs is often attributed to synergistic effects where the presence of more than one metal in the catalyst improves its catalytic properties beyond the sum of the properties of the different metals tested separately. In the case of Au-Cu NPs it has been shown that the addition of Cu to Au has a promoting effect on different reactions such as carbon monoxide oxidation reaction,[14,15,16,17,18] PROX reaction (preferential CO oxidation in the presence of hydrogen),[15,19,20,21,22,23,24] propene epoxidation,[25,26] selective hydrogenation of butadiene[27,28] and cinnamaldehyde[29]. More recently Au-Cu NPs have shown a promising efficiency in the challenging fields of fuel production from carbon dioxide[30,31,32,33] and of wastewater treatment,[34] combining the catalytic properties of the NPs and the photo-catalytic properties of a $TiO_2$ support. However, synergistic effects alone do not explain all the discrepancies between monometallic and bimetallic NPs regarding the catalytic performance and it is often difficult to determine the origin of this difference.[35]

Several studies have shown that in addition to the size, the structure of the NPs and the ability of Au and Cu to form an alloy, and/or Cu to form an oxide, have a crucial influence on the activity of the NPs. For Au-Cu NPs deposited on silica gel[15] and SBA-15,[14] X-Ray Diffraction (XRD) spectra measured after reduction under $H_2$ at 820K for a series of composition ranging from pure Au to pure Cu showed the formation of alloyed particles of about 3 – 4 nm in diameter. In the case of $TiO_2$ supported Au-Cu NPs, the presence of a Cu oxide layer sandwiched between the substrate and the Au NPs has been shown to improve the activity of the NPs.[16] It was thus suggested that the synergistic effect occurs between Au and Cu oxide.[36] Bauer et al. have compared the activity of Au-$CuO_x$/$SiO_2$ NPs to Pt/$Al_2O_3$ NPs for the CO oxidation below 150°C and measured a significant improvement for Au-$CuO_x$/$SiO_2$ as well as a stabilisation attributed to an 'anchoring' effect of the NPs over the silica support.[37] Delannoy et al. reached a similar conclusion for selective oxidation of butadiene by Au-Cu NPs on $TiO_2$.[28] An initial activation phase of the catalyst was observed during the reaction which was tentatively attributed to the segregation of Cu at the surface of the NPs. The influence of catalyst pre-treatments such as reduction and calcination have recently been investigated by Liao et al. for the PROX reaction for Au-Cu on $CeO_2$[23] and on $Al_2O_3$ NPs.[24] In both cases, a reduction pre-treatment leading to alloyed NPs was shown to improve the CO conversion and the stability of the NPs in the presence of reactants. Liquid phase selective hydrogenation of cinnamaldehyde by Au-Cu/$CeO_2$ NPs was investigated in mild conditions (70°C, 1 atm).[29] The structure of the NPs after the reduction pre-treatment was investigated using HAADF-STEM and was shown to form alloyed NPs with no visible segregation.

This highlights the growing interest for *in-situ* experiments where the structure of the NPs is continuously monitored as the system goes through the whole catalytic process from activation up to poisoning. Grazing incidence x-ray diffraction (GIXD) using synchrotron radiation is an excellent technique to carry out this type of investigation as the low incidence angle allows to be extremely sensitive to the surface and the high photon energy used (5 – 25 keV) allows the beam to penetrate dense reactive phases (gas or liquid) with limited absorption. In this study, *in-situ* GIXD was used to fully characterize the structure of Au-Cu NPs deposited onto $TiO_2$ (110) single crystals. The evolution of the structure of each type of Au-Cu NPs was monitored *in-situ* during subsequent and simultaneous exposure to oxygen and carbon monoxide. The structural characterization was then complemented with x-ray photoemission spectroscopy (XPS) investigation that corroborates GIXD findings and unveils the diffusion mechanisms leading to the formation of Cu oxide that may explain the enhanced activity of bimetallic Au-Cu NPs.

**II) Experiment**

All experiments were performed on UHV set-ups with a base pressure of a few $1 \times 10^{-10}$ mbar. Rutile $TiO_2$ (110) single crystals (from MaTecK GmbH) were cleaned by several cycles of Ar sputtering ($P_{Ar}$ = $7 \times 10^{-5}$ mbar, 20 minutes, E = 1 keV) and annealing under UHV (10 minutes at 1000 K), leading to slightly reduced samples. Au and Cu were evaporated *in situ* under UHV either from e-beam evaporators or from effusion cells, using high purity Au and Cu (Alfa Aesar). The amounts of Au and Cu evaporated were measured during the experiments with a flux monitor in the case of the e-beam evaporators and with a quartz microbalance in the case of the effusion cells. During each experiment a calibration sample was realized by depositing a large amount of Au and Cu (typically 2 ML of each) on a $TiO_2$ sample. The quantity of Au or Cu present on the calibration samples was precisely determined using Rutherford backscattering spectroscopy (RBS) performed with a van de Graaff accelerator at INSP. An excellent agreement was found between doses estimated using the flux monitor or the quartz microbalance and the values obtained using RBS. Ultrapure $O_2$ and CO were introduced in the chambers using precision leak valves.

GIXD experiments were performed at the SixS beamline of SOLEIL synchrotron and at the BM32 beamline of ESRF synchrotron. GIXD measurements were performed at an energy of 15 keV and a grazing incidence angle of 0.2°. The x-ray diffraction data presented in this study are indexed according to the surface unit cell of $TiO_2$ (110) described by the following parameters: **a** = 0.296 nm, **b** = **c** = 0.648 nm, and α = β = γ = 90°. h, k, and l Miller indices are used for indexing reflections in the reciprocal space. As a matter of clarity some of the data have been plotted against the modulus of the momentum transfer corresponding to $q = 2\pi \cdot \left(\frac{h^2}{a^2} + \frac{k^2}{b^2} + \frac{l^2}{c^2}\right)^{1/2}$. The NPs lattice parameters were determined for each peak using the centre of a Gaussian fit. The size of as-prepared NPs were determined after the synthesis using the Scherrer equation[38] and a height/diameter parameter for the NPs of 0.89, obtained from previous investigations.[42] However, due to the experimental set-up utilised (point detector and detector slits) and the fact that the aspect ratio of the crystalline domain of the NPs was evolving during the experiment, no adequate parameter was found to input in the Scherrer formula to satisfactorily represent the evolution of the size of the NPs.

XPS experiments were performed at the TEMPO beamline of SOLEIL synchrotron. Au 4f and Cu 3p core levels were recorded in the same kinetic energy window using normal emission of the photoelectrons and photon energies of 250 eV, 700 eV and 1050 eV with an overall resolution E/$\Delta$E of about 5000. Intensities were normalized to the incoming photon flux recorded on a gold mesh located prior to the analysis chamber. Shirley background has been removed from all XPS spectra. The binding energy (BE) has been calibrated with respect to the Fermi energy measured on the Ta stripes holding the sample.

**III) GIXD results**

   **a) NPs structure under UHV**

Hereafter, we present the results obtained under UHV on a reference sample composed of $Au_{52}Cu_{48}$ NPs of about 2 nm in size grown on a rutile $TiO_2$(110) substrate. Those NPs have been obtained by the successive evaporations of $9.5 \times 10^{14}$ at.cm$^{-2}$ of Au at 570K and $8.8 \times 10^{14}$ at.cm$^{-2}$ of Cu at room temperature. This procedure has previously been shown to lead to the formation of bimetallic particles.[42]

Figure 1(a) shows a (h, k, l=0.05) map of the intensity scattered by the sample in the reciprocal space. Intense sharp peaks (red/black) matching integer values of *h* and *k*, correspond to Bragg reflections of the $TiO_2$ substrate. The broad spots at non-integer values of *h* and *k* and the ring centred at the origin are due to the scattering by the NPs. The broad spots are attributed to crystalline NPs that have grown on the $TiO_2$(110) substrate with specific epitaxial relationships detailed in figure 1(b). The ring corresponds to randomly oriented NPs. In the following, the first type of NPs will be referred as epitaxial NPs and the second type as randomly oriented NPs. Figure 1(b) is a schematic representation of the different features present in figure 1(a). The black spots correspond to the substrate Bragg peaks, blue, red and green spots to diffraction peaks of oriented NPs and the purple arc to the signal obtained from randomly oriented NPs.

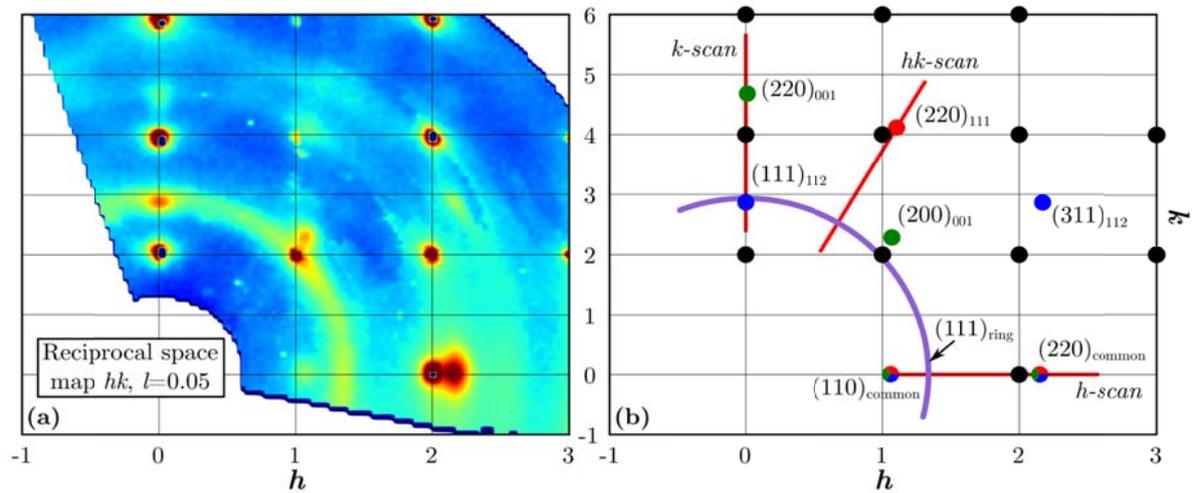

FIG 1: (a) 2D (h, k, l = 0.05) map of the x-ray intensity scattered by a sample composed of 2 nm $Au_{52}Cu_{48}$ NPs evaporated on rutile $TiO_2$(110). (b) schematic representation of (a). Black dots: $TiO_2$ Bragg reflections, red, green and blue dots: reflections for epitaxial NPs, purple ring: intensity scattered by randomly oriented NPs. The three red lines marked 'k-scan', 'hk-scan' and 'h-scan' show the direction of the reference scans drawn in figure 2.

The structure and epitaxial relationship of oriented NPs were determined by indexing the spots shown on the map 1(a). As in previous studies, the epitaxial relationships were determined by indexing all the spots. All diffraction spots match a slightly distorted cubic structure. The presence of a (110) peak demonstrates a chemical order of the epitaxial NPs, as compared with a chemically disordered *fcc* structure for which this peak is extinct. These observations are in good agreement with the slightly quadratic $L1_0$ structure (c/a = 0.935) that is the stable bulk structure for this composition. For all the other deposits studied, corresponding to different compositions, the (110) peak was not present. However, we always found a similar matching with a slightly distorted cubic structure. A small distortion with respect to a cubic structure was also found for pure Au NPs prior to Cu evaporation. For the sake of simplicity, the spots have been indexed with respect to a cubic structure and the specific lattice constants for each spots were measured throughout the experiments.

Even if they correspond to the same cubic structure, all diffraction spots found do not belong to a single epitaxial relationship. Three epitaxial relationships, already reported in previous studies[39,40,41,42] for Au NPs are present: $(111)_{NPs}//(110)_{TiO2}$, $(001)_{NPs}//(110)_{TiO2}$ and $(112)_{NPs}//(110)_{TiO2}$. For each of these epitaxial relationships, the $[1-10]_{NPs}//[001]_{TiO2}$ condition is also satisfied, associated with an intense $(220)_{common}$ peak. For the sake of clarity, the reflections measured are indexed as $(hkl)_{(mnp)}$ where (hkl) is a given Bragg reflection which arises from NPs whose (mnp) plane is parallel to the $(110)_{TiO2}$ surface. Hence, using GIXD, it is possible to follow the evolution of each specific type of NPs upon gas exposure by collecting the intensity scattered in the reciprocal space (red lines in figure 1(b)). The NPs lattice parameter measured from the $(111)_{112}$ spot is 0.382 nm, it corresponds to a $Au_{47}Cu_{53}$ alloy according to the Vegard's law. This is consistent with the average sample composition ($Au_{52}Cu_{48}$) and indicates that the successive evaporation of Au and Cu forms bimetallic alloyed NPs of the desired composition.

### b) CO exposure followed by CO + $O_2$ exposure

In a second experiment, $Au_{38}Cu_{62}$ NPs obtained by the successive deposition of $8.6 \times 10^{14}$ at.cm$^{-2}$ of Au and $1.4 \times 10^{15}$ at.cm$^{-2}$ of Cu on a $TiO_2$ substrate at 415 K have been exposed, at room temperature, to CO and to a CO+$O_2$ mixture. In order to follow the evolution of the structure of each type of NPs the *h-, k- and hk-* reference scans corresponding to the red lines shown in figure 1(b) have been acquired repeatedly before and during gas exposure.

Figure 2 shows the comparison between the same *h-, k-* and *hk-*scans acquired in UHV after growth (green triangles in figure 2(a), (b) and (c) respectively), after a first CO exposure ($P_{CO} = 1 \times 10^{-6}$ mbar; red squares) and after a subsequent exposure to a CO + $O_2$ (1:1) mixture ($P_{CO+O2} = 2 \times 10^{-6}$ mbar; blue circles). The structure of the as-prepared Au-Cu NPs is determined from the reference scans realized under UHV condition (green triangles). The peak around 28.7 nm$^{-1}$, which is present on all the scans, corresponds to the (111) *ring* feature previously ascribed to randomly oriented NPs on the substrate. In the case of UHV, it is associated with a lattice parameter of 0.380 nm indicating that NPs are alloyed. The presence of a small $(220)_{common}$ peak on the *h*-scan around 45.6 nm$^{-1}$ shows the presence of epitaxial NPs. However, the absence of the $(220)_{001}$ peak in the *k*-scan and the weak intensity of the $(220)_{111}$ peak in the *hk*-scan as compared to the *ring* intensity, suggest that most of the NPs are randomly oriented. A lattice parameter of 0.390 nm was extracted from the $(220)_{common}$ peak showing that epitaxial NPs are also alloyed. From the width of the diffraction spots, we estimate that the randomly-oriented particles have a mean size of 1.8 nm, whereas (112) epitaxial nanoparticles have a mean size of 2.0 nm. The difference in lattice parameter between epitaxial and randomly oriented NPs could either be attributed to a slight difference in composition or to a different of strain between the two types of NPs. We have previously shown that for small Au particles, a decrease of the lattice constant is observed when the size decays. [42,43] In the present case, we do not observe significant differences between the value of the position of the (111) peak found for randomly-oriented particles and for (112) oriented particles. The difference with the position of the $(220)_{common}$ peak could thus more probably reflect a lattice distortion of the cubic lattice. However the proximity of the $(220)_{common}$ peak and the $TiO_2$ Bragg peak may influence the value of the fit.

The introduction of $10^{-6}$ mbar of CO has practically no influence on any of the peaks measured along the *h-, k- and hk-* directions as can be seen in figure 2 (red squares). However, the addition of $10^{-6}$ mbar of $O_2$ leading to a total pressure of $2 \times 10^{-6}$ mbar of CO + $O_2$ in (1:1) proportion results in a drastic change of the structure for both epitaxial and randomly oriented NPs (blue circles in figure 2).

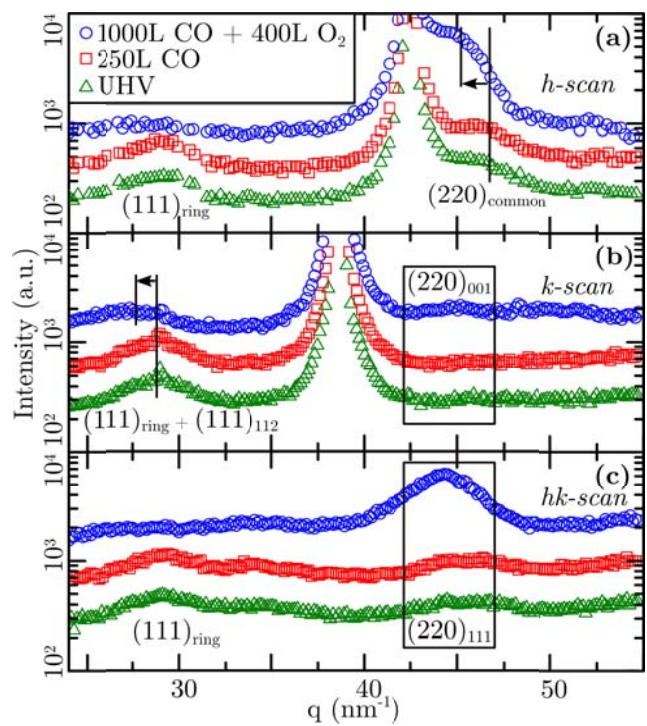

*FIG 2. Diffracted intensity for $Au_{38}Cu_{62}/TiO_2$ NPs, as a function of momentum transfer, along different directions of the reciprocal space, the curves have been shifted for clarity. (a) for k=0, l=0.05. (b) for h=0, l=0.05. (c), for k=3.79h, l=0.05. Green triangles: as prepared NPs. Red squares: during $1 \times 10^{-6}$ mbar CO exposure. Blue circles: during $1 \times 10^{-6}$ mbar CO + $1 \times 10^{-6}$ mbar $O_2$ exposure. The dashed rectangles indicate the regions where the (220) reflections of epitaxial NPs are expected. The plain vertical lines show the shift of the peaks after the introduction of oxygen.*

To follow precisely the sample evolution, all the peaks presented in the reference scans have been fitted with a Voigt profile. The lattice parameter and integrated intensity associated to the best fit for each scan are shown in figure 3 (a) and (b) respectively, as a function of CO and $O_2$ dose. The parameters of the $(111)_{ring}$ and $(220)_{common}$ peaks (green triangles and red squares respectively) were fitted from the *h-scans*. The parameters of the $(111)_{112}$ (blue circles) were fitted from the *k-scans* and the parameters of the $(220)_{111}$ (purple crosses) from the *hk-scans*.

As already mentioned, no evolution is observed after introduction of CO, but important changes of the lattice parameters occurred immediately after the introduction of $O_2$ (first dashed line from the left) and on the integrated intensity after exposure to 100 L of $O_2$ (second dashed line from the left). Eventually, the lattice parameter shifts from 0.386 nm for (111) reflections and 0.390 nm for the (220)$_{common}$ reflection to about 0.402 nm corresponding approximately to the lattice parameter found for pure Au NPs of similar size.[42] At this point, no evolution is seen upon further exposure under the same conditions (10$^{-6}$ $O_2$). The most remarkable point is that the changes occurring to the integrated intensity of the peaks depend on the type of NPs. The signal from the *ring* feature drops to a point where after 300L of $O_2$, no peaks can be fitted due to a low signal/noise ratio as seen in figure 2(a) and (c). A shifted (111) reflection remains on the figure 2(b) as it corresponds to the (111)$_{112}$ reflection of epitaxial NPs. Interestingly, a specific epitaxial relationship is favoured, as shown by the gain in intensity. The (220)$_{111}$ contribution (figure 2 (c) and purple crosses figure 3 (b)), which could not be fitted initially due to a low signal/noise ratio, becomes the most intense peak with the exception of the (220)$_{common}$ peak. In the meantime, no drastic gain in intensity is observed for the (111)$_{112}$ reflection and no peak can be fitted for the (220)$_{001}$ reflection as seen in the dashed rectangle in figure 2 (b).

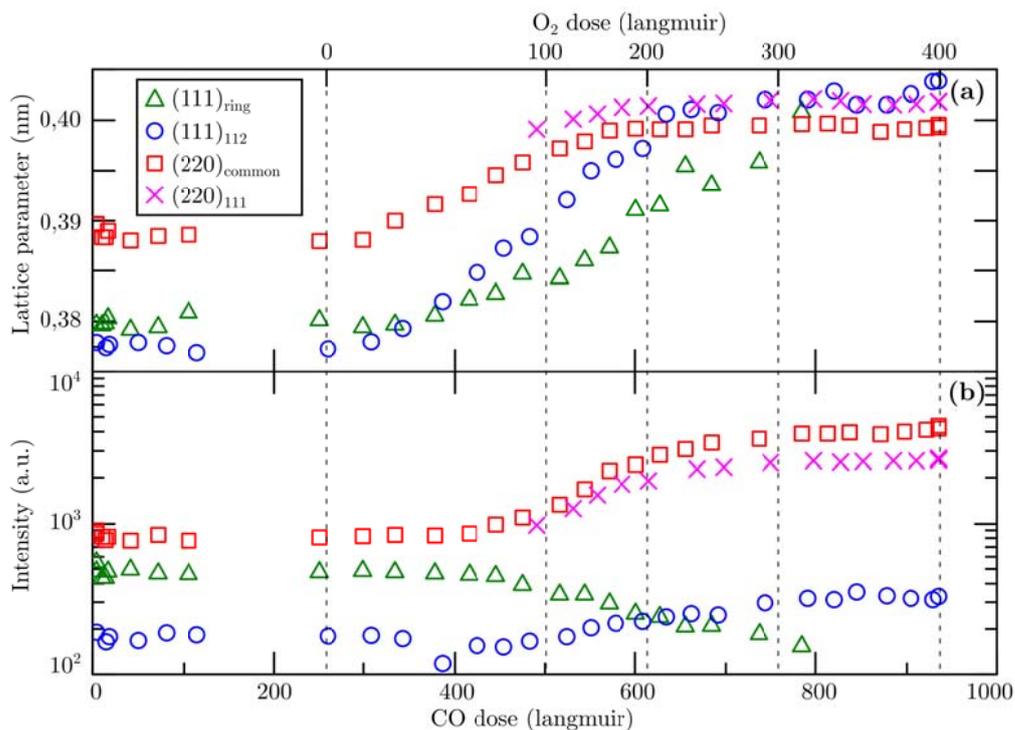

*Fig 3: Evolution of the lattice parameter (a) and of the intensity (b) associated with diffraction peaks in the reference scans (h-scan, hk-scan or k-scan). Green triangles: (111)$_{ring}$, blue dots: (111)$_{112}$, red squares: (220)$_{common}$, purple crosses: (220)$_{111}$. The intensity associated with the (111)$_{ring}$ peak drops below the background level after 300 L of $O_2$ exposure.*

After this exposure to CO and O$_2$, the gas was evacuated from the chamber and the sample was kept under UHV for several hours. No evolution of the diffraction signal was observed during this step. The NPs have then been annealed progressively from room temperature up to 470 K under UHV. During this process, the evolution of the lattice parameter obtained from the different peak positions is represented in figure 4. Up to 420 K, no significant variation can be observed: the lattice parameter remains close to the one of pure Au. At a temperature of 470 K, a clear decrease of the lattice parameter of the NPs is measured which could be associated with the partial recovering of the initial NP state, alloyed NPs. However, for the annealing conditions used here, the initial lattice parameter is not recovered.

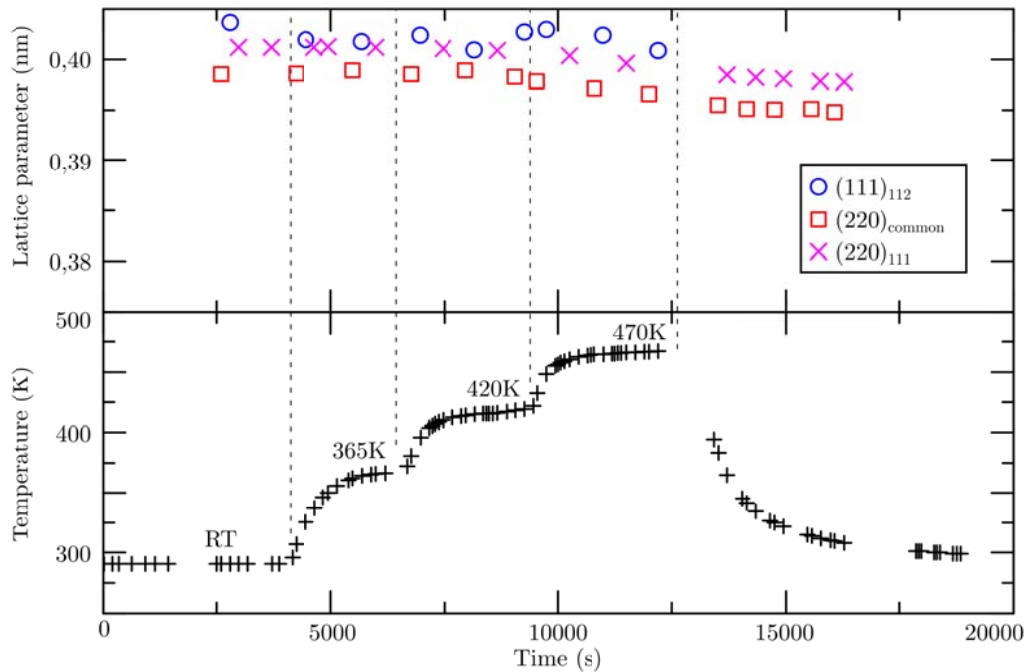

*Figure 4: Evolution of the lattice parameter measured for the (111)$_{112}$, (220)$_{common}$ and (220)$_{111}$ peaks (respectively in blue, red and purple) as a function of time as the temperature of the substrate is increased progressively to 365K, 420K and 470K.*

### c) O$_2$ exposure followed by CO + O$_2$ exposure

A third sample has been prepared at room temperature with the co-evaporation of 8.3 × 10$^{14}$ at.cm$^{-2}$ of Au and 8.8 × 10$^{14}$ at.cm$^{-2}$ of Cu, which should lead to Au$_{49}$Cu$_{51}$ composition. The NPs obtained have a lattice parameter equal to 0.384 nm, which corresponds to a composition close to Au$_{51}$Cu$_{49}$. For this sample, particles were mainly epitaxial following (111) and (112) epitaxies and had an estimated size after growth of around 1.9 nm. The signal associated with randomly oriented particles corresponded to small NPs with a mean size of around 1 nm.

As compared with sample 2, sample 3 has thus a higher Au content. The figure 5 shows the evolution as a function of time of the lattice parameter of epitaxial NPs measured from the (220)$_{common}$ peak during the different steps of gas exposure and annealing (Figure 5).

During the step 1, the sample was exposed to approximately $1 \times 10^{-6}$ mbar $O_2$ (purple circles) at room temperature. A clear shift of the lattice parameter from 0.384 to 0.392 nm is visible. No significant variation of the relative intensity between the different peaks could be observed during this oxygen exposure. After $O_2$ exposure, the sample has been maintained in UHV during step 2 at room temperature. No significant evolution of the diffraction signal has been observed during this period, which indicates that the structure formed during oxygen exposure is stable in UHV. The sample was then exposed during step 3 to approximately $10^{-6}$ mbar of CO (blue triangles), for temperature increasing from 300 K up to 380 K, and no evolution of the lattice parameter was observed, showing that the structure formed during oxygen exposure is also stable in the presence of CO.

However during step 4, annealing the sample at 430 K under CO induces a slow decrease of the lattice parameter. This threshold is indicated by the third vertical dashed line in figure 5. After annealing up to 500 K, the lattice parameter of the NPs reaches a value of 0.387 nm, close to the one of as prepared NPs.

During step 5 the sample was cooled down to room temperature under UHV and then re-exposed to $O_2$ during step 6. The lattice parameter of the NPs increases again to higher values

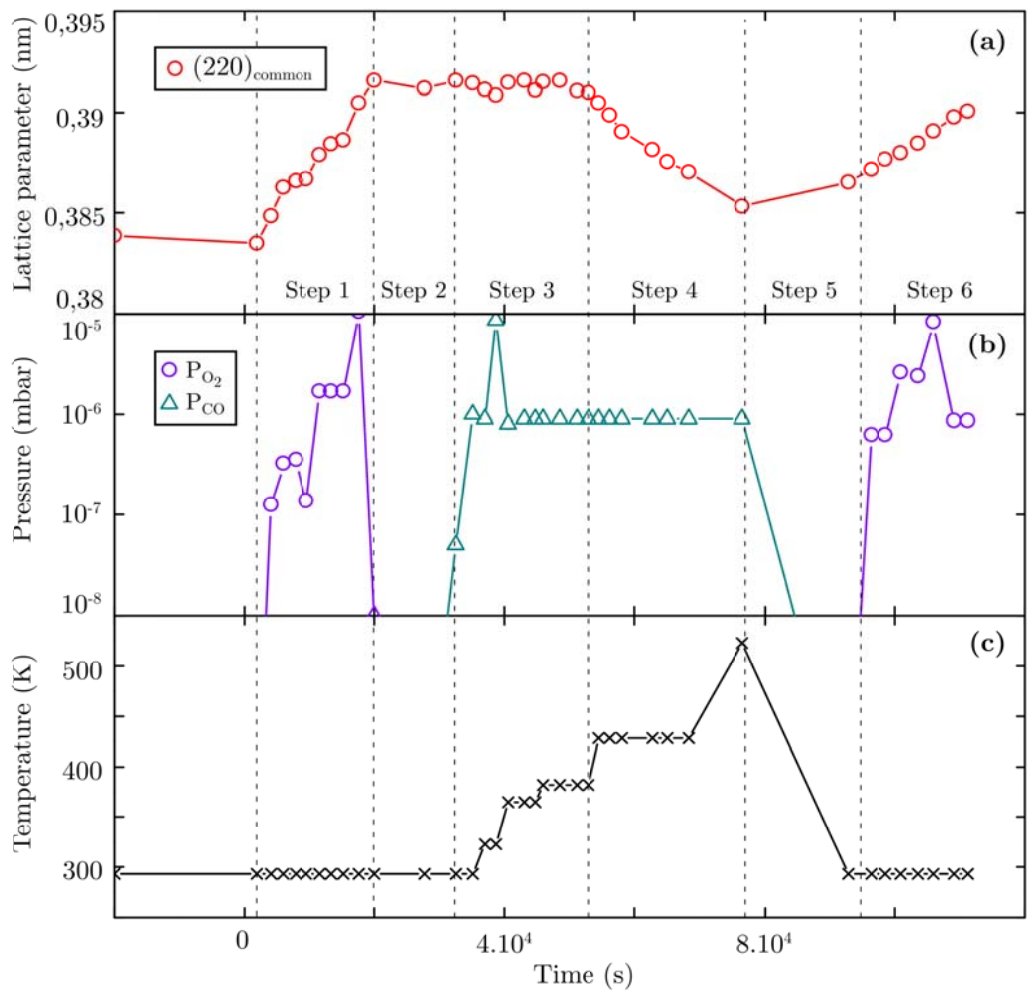

showing that a reversible process is occurring between oxygen introduction and annealing.

*Figure 5: Time of evolution of: a) lattice parameter of $Au_{51}$-$Cu_{49}$ NPs measured from the position of the $(220)_{common}$ peak, (b) $O_2$ (blue) and CO (purple) pressures and (c), temperature.*

### IV) XPS results

Two other samples have been prepared for XPS experiments with compositions of $Au_{27}Cu_{73}$ and $Au_{53}Cu_{47}$. The XPS spectra of Au 4f and Cu 3p core-levels, recorded on these two samples for a photon energy of 250 eV, are displayed in the figure 6. The top spectra corresponds to the $Au_{27}Cu_{73}$ sample, whose composition has been obtained from Rutherford Backscattering Spectroscopy (RBS) measurement after the removal of the sample from the XPS chamber. The Au 4f and Cu 3p intensities have been extracted by the deconvolution of each component using Voigt functions. Taking into account the cross section at 250 eV of photon energy[44] and the mean free path of electrons in copper and gold,[45] the quantitative analysis of Au 4f and Cu 3p leads to a composition of $Au_{24}Cu_{76}$, close to the one obtained using RBS. This validates the procedure used to determine the atomic fraction of Cu and Au by XPS. The bottom spectra of the figure 6 have been recorded on a second sample, with $Au_{53}Cu_{47}$ composition, as given by the flux of evaporation. This composition is confirmed by XPS analysis at 250 eV photon energy, giving $Au_{51}Cu_{49}$.

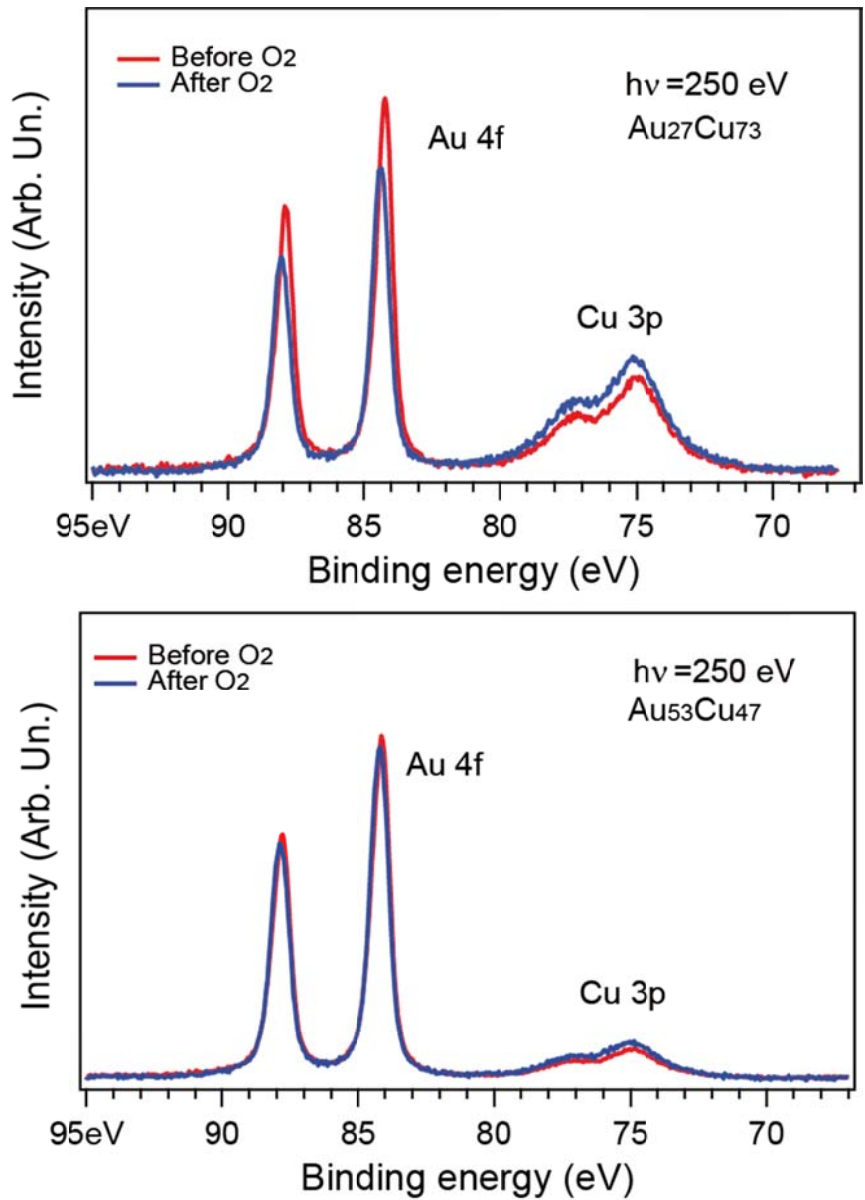

*Figure 6: Au 4f and Cu 3p XPS spectra recorded on the $Au_{27}Cu_{73}$ NP (top) and $Au_{53}Cu_{47}$ NPs (bottom) at a photon energy of 250 eV before and after exposition to molecular oxygen.*

When the particles are exposed to up to $2 \times 10^{-5}$ mbar of molecular oxygen at room temperature, a decrease of the 4f peak and an increase of the Cu 3p peak is observed for both samples. In the case of the copper rich nanoparticles, the change in intensity is more pronounced and a slight energy shift to higher binding energy of about 50 meV of the gold and copper peaks is observed, which is in the limits of the energy resolution. Because the NPs are deposited on a $TiO_2$ substrate, the O 1s core-level (not shown here) is dominated by the substrate contribution and cannot give information on the possible oxidation of the NPs. Within the energy resolution, no chemical shift could be seen on the Auger Cu LMM peaks and on the Cu2p peaks. Further exposure to a mixture of CO+$O_2$ or to pure CO (up to 2 × $10^{-5}$ mbar) does not lead to noticeable changes in the XPS spectra.

Figure 7 displays the evolution of the Cu fraction as a function of the incident photon energy, for as prepared NPs (blue) and after $O_2$ exposure (red). The value has been obtained from the tabulated values of the XPS cross-sections and electron mean free path. Due to the uncertainties related to these values, it is not possible to relate the energy dependence of the values found to a clear segregation for as-prepared NPs. After oxidation, the Cu atomic fraction increases. The strongest variations are observed for low photon energy, for which the escaping electrons originate from the near surface region of the NPs. In this region, an increase of the Cu atomic fraction is observed after $O_2$ exposure. Thus, both GIXD and XPS results are in agreement with a segregation phenomenon where a demixing of Au and Cu occurs in the NPs, leading to the formation of a Au-rich core, and a Cu-rich surface.

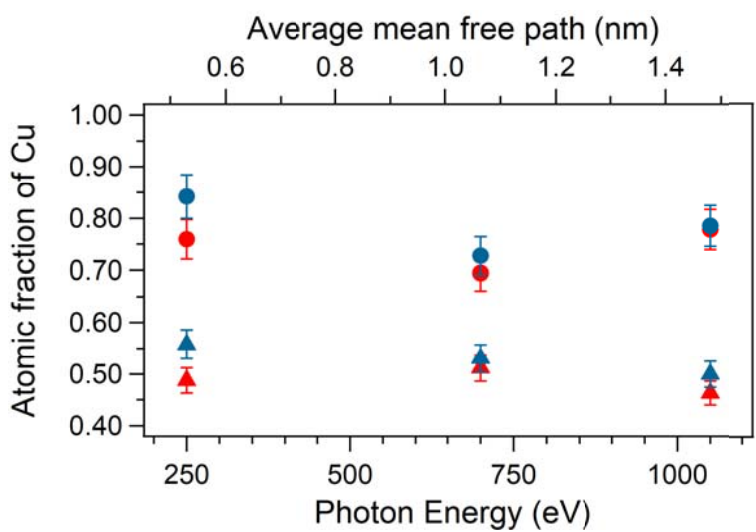

*Figure 7: Atomic fraction of copper vs photon energy, obtained from XPS measurements, for $Au_{53}Cu_{47}$ (triangle) and $Au_{27}Cu_{73}$ (dot) before (red) and after (blue) oxidation.*

## V) Discussion
### a) Cu surface segregation phenomenon and formation of a copper oxide shell

For both NPs samples presented in section III.b) and c), a shift of the lattice parameter towards higher values is observed immediately after the introduction of molecular oxygen, independently of the presence of CO. This indicates the formation of a Au-rich phase. The photoemission experiments show that samples of close composition exposed to $O_2$ display an increase of the Cu 3p peak intensity and a decay of the Au 4f peak intensity. In addition, GIXD results show the absence of any peak related to crystalline copper (with a lattice parameter close to the one of bulk Cu = 0.361 nm).

From XPS, we can exclude the formation of a large amount of a CuO oxide, which would have led to a chemical shift larger than 1 eV on the Cu 2p and Cu 3p peaks. The small chemical shifts observed with XPS may be associated with the formation of a shell of $Cu_2O$ or to chemisorbed oxygen. However, the absence of significant shift on the Auger peak indicate that most Cu atoms remain in a metallic state. This high stability towards oxidation could be due to the low $O_2$ partial pressure used in the present experiments. It has been shown using Scanning Tunneling Microscopy (STM) that the morphology of Au-Cu NPs of about 1 nm in size remains stable during the introduction of $10^{-6}$ mbar of oxygen if they incorporate more than 20% of Au.[46] The fact that the individual volume of the NPs remains the same means that the Cu does not diffuse away from the particles but forms a core-shell structure with a Au-rich core and a $Cu_2O$ shell too poorly crystallised to scatter significant intensity.

Such a phenomenon was studied using XPS and density functional theory (DFT) for Au-rich polycrystalline surfaces by Völker and co-workers[47] and for $Cu_3Au(111)$ by Tsuda and co-workers.[48,49] In both cases the XPS/DFT interplay shows that the alloyed surfaces under UHV are slightly enriched in Au and that a clear inversion occurs during the introduction of oxygen with Cu diffusing to the outermost layers of the surface. More recently, Dhifallah and co-workers have suggested using first-principles calculation that the Cu segregation may also depend on the surface orientation.[50]

**b) Thermal stability of the core/shell structure**

In each of the GIXD experiments presented in figure 4 and 5 the gas has been pumped down and the system maintained under UHV at room temperature to investigate a possible diffusion of the Cu atoms back into the Au enriched core of the NP. Indeed, it has been shown that Cu atoms deposited onto ≈ 2 nm Au NPs at room temperature diffuse into them to form an alloy.[42] The fact that such diffusion could not be observed reinforce the interpretation made previously that the copper forms an oxide shell when exposed to oxygen. This shell is stable at room temperature, under UHV or when exposed to $10^{-6}$ mbar of CO (see figure 5). This interpretation is in agreement with the conclusion of Sandoval and co-workers[18] regarding the formation of Cu oxide but in the present case, XPS demonstrate that the amorphous Cu oxide cannot be 'sandwiched' between the substrate and Au clusters. Moreover, in such a configuration, the epitaxial relationship between the NPs and the substrate would have been lost, which is not the case here.

The Cu oxide shell can be reduced by annealing the sample as shown in figure 4 and 5 at a temperature above 430 K. The decrease of the lattice parameter is therefore interpreted as Cu atoms being progressively released from the shell and forming a Au-Cu alloy within the NP. As expected from this scenario, the re-introduction of oxygen at room temperature after annealing regenerates the Cu oxide shell through a segregation phenomenon as seen in figure 5 during the second introduction of oxygen.

### c) Gas-induced selective re-orientation/crystallisation of the NPs

The disappearance of the ring feature upon introduction of a mix of CO + $O_2$ implies a deep and global reorganisation of individual NPs. The sample presented in section b) for which the re-orientation was observed, was synthesized under UHV on a substrate at 415 K to improve the diffusion of individual Au and Cu atoms and obtain larger NPs. Despite this procedure, the signal scattered by randomly oriented NPs was the majority. In other experiments, we observed that sample annealing up to 670 K in UHV does not change the intensity of the ring and does not improve the epitaxial relationships of the NPs (not shown here).

Two interpretations can be made to explain the total disappearance of randomly oriented NPs in the presence of the gas mixture. i), a NP to NP diffusion mechanism occurs where randomly oriented NPs, which may also be the smallest, sinter. Au and Cu atoms diffuse from these small NPs toward larger, epitaxial and more stable NPs, resulting in an increase of the corresponding diffracted intensity. ii) an intra-atomic mobility triggers the NP re-crystallisation, favouring a $(111)_{NPs}//(110)_{TiO2}$ epitaxy. The second hypothesis is more likely because a ripening mechanism cannot explain that a specific epitaxial relationship is favoured. The irreversibility of the disappearance of the ring feature, even after annealing, is in agreement with both previous interpretations. The disappearance of randomly oriented NPs is probably triggered by the fact that the chemical potential is higher for these NPs than for epitaxial ones. This could be due to a weaker interfacial bonding or to a smaller size (1.8 nm instead of 2.0 nm). On the one hand, the adhesion energy of a Au particle on $TiO_2(110)$ has been shown to depend on its geometry.[51] On the other hand, from the surface energy of Au and Cu[52], a difference in chemical potential of 0.05 eV/at is expected between 1.8 nm and 2.0 nm size $Au_{50}Cu_{50}$ particles.

At this point, one has to determine whether the disappearance of randomly oriented NPs is induced by the mixture of gas or if such effect could be seen for pure oxygen exposure. The segregation mechanism has been shown to occur in the presence of oxygen only (see figure 5) and if, as stated above, the re-orientation mechanism was a consequence of Au and Cu atoms segregating it would happen in these conditions as well. However, if this phenomenon has been observed under a CO+$O_2$ mixture on the $Au_{38}Cu_{62}$ sample, it has not been measured on the $Au_{49}Cu_{51}$ sample under pure $O_2$. Nevertheless, in that case, a higher Au content may enhance the NP stability. We have indeed previously shown by STM that $Au_xCu_{1-x}$ NPs were stable under $10^{-5}$ mbar $O_2$ for x>0.2.[46]

It is worth to note that the atomic mobility needed for the reorientation of the NPs is greater than the one required for Cu segregation to the surface, since it implies that all atoms, including those at the interface with the substrate, are involved. Au-Cu/$TiO_2$ NPs have also been shown to catalyse the carbon monoxide oxidation at room temperature and above.[18] Thus, considering that the reaction is exothermic (2.93 eV per $CO_2$ molecules), one can suggest that the energy released by the reaction, occurring here at room temperature, leads to an enhanced atomic mobility, and then to a *gas-induced selective re-orientation* of the NPs at the pressure at which this study has been performed.

**VI) Conclusion**

The combined use of GIXD and XPS allows us to comprehend the atomic diffusion and oxidation mechanisms at play in the case of 2 nm Au-Cu NPs exposed to oxygen and carbon monoxide. When exposed to $10^{-6}$ mbar of oxygen the Cu atoms, initially alloyed with Au, segregate to the surface of the NPs to form a Cu oxide shell stable in UHV at room temperature and reducible by annealing the sample over 430K. In the presence of a CO + $O_2$ mixture, this segregation phenomenon triggers a second phenomenon referred here as *gas-induced selective re-orientation* where randomly oriented NPs use the mobility induced by the segregation to re-orient their Au-rich core following the specific epitaxial relationship $(111)_{NPs}//(110)_{TiO2}$, $[1-10]_{NPs}//[001]_{TiO2}$. It remains unsure whether the presence of carbon monoxide is mandatory to trigger the re-orientation mechanism. After reduction, the alloyed NPs keep their epitaxial relationship making this procedure an efficient way to selectively orient Au-Cu NPs on a rutile $TiO_2(110)$ substrate.

**Acknowledgements**

Axel Wilson has received funding from the European Union's Horizon 2020 research and innovation programme under the Marie Skłodowska-Curie grant agreement (GA) No 665593 awarded to the Science and Technology Facilities Council. Experiments were performed on the SixS and TEMPO beamlines at SOLEIL Synchrotron, France and on the BM32 beamline at the ESRF, France.